# Recyclable Flame-retardant Epoxy Composites Based on Disulfide Bonds: Flammability and Recyclability

Xiaolu Li, [a,b] Jing Zhang, [a,b] Lu Zhang, [a,c] Alaitz Ruiz de Luzuriaga, [d] Alaitz Rekondo, [d] De-Yi Wang [a] *

[a.] IMDEA Materials Institute, C/Eric Kandel, 2, 28906 Getafe, Madrid, Spain

[b.] E.T.S. de Ingenieros de Caminos, Universidad Politécnica de Madrid, Calle Profesor Aranguren 3, 28040 Madrid, Spain

[c.] College of Mechanical and Electrical Engineering, Central South University, 410083 Changsha, Hunan, P.R. China

[d.] CIDETEC, Basque Research and Technology Alliance (BRTA), Paseo Miramón, 196, 20014 Donostia-San Sebastián, Spain

Email: deyi.wang@imdea.org

## Abstract

A series of recyclable epoxy resin (REP) composites were successfully prepared in this study, aiming to improve flame retardancy while keep recyclability and mechanical properties of epoxy. The 9,10-Dihydro-9-oxa-10-phosphaphenanthrene 10-Oxide (DOPO) with high reactivity on the P-H bond was used as highly-efficient flame retardant in this epoxy system. Compared with pure epoxy resin with a limited oxygen index (LOI) of 21.7 % and no rating in vertical burning test (UL-94), recyclable epoxy resin only with 3 *wt*% DOPO acquired a LOI value of 27.8% and passed V-0 rating. Based on the results of cone calorimeter test (CCT), REP composites containing 4 *wt*% DOPO had excellent self-extinguish, showing the 43.5% drop of the peak heat release rate (pHRR). This was attributed to the charring ability in condensed phase and free radical scavenging in gas phase caused by traditional flame retardant DOPO, which can play a positive role in inhibiting combustion when burning. In addition, both scanning electron microscopy (SEM) and Raman results also exhibited outstanding flame retardancy. Importantly, superb dispersion between epoxy and DOPO provided fine transparency and mechanical properties of REP composites. Additionally, REP composites with special exchangeable disulfide crosslinks had fantastic recyclability. REP composites possessing excellent flame retardancy, recyclability and transparency pave the way to broaden application of thermoset epoxy materials.

## 1. Introduction

Epoxy resin (EP) has proven to be one of the most prevalent thermosetting resins on account of convenient curing processes, superior mechanical property, low shrinkage and strong designability[1, 2]. So far, EP materials have been used in industrial appliances of aerospace, construction, coating, etc. However, once thermoset materials are damaged, they are difficult to dissolve and reshaped again because of crosslinking structure formed between segments, which will lead to a waste of resources and cost increase. This limits some applications indirectly. In addition, thermoset EP are flammable materials. Thus, recyclability and flame retardancy of EP materials are still challenging.

For improving repairability of epoxy, dispersing the thermoplastic healing agent into EP to form shape memory matrix is one method[3, 4]. Another common strategy is adding microencapsulated healing agent and catalyst[5-10]. Once thermoset materials are destroyed, healing agent is released to start self-healing process. Recently, the introduction of dynamic chemical bonds has been the novel way to promote repairability and recyclability, accompanying by the emergence of some new concepts "vitrimers, dynamers and covalent adaptable networks (CANs)". Among them, the concept of "vitrimers" puts forward by Leibler and co-workers[11] and vitrimer materials developed according to the dynamic exchange reaction are the first recyclable thermosets. The popular dynamic exchange reactions include carboxylate transesterification[12], transcarbamoylation[13], siloxane silanol exchange[14], disulfide exchange[15-18] and imine amine exchange[19]. However, the described vitrimers, still have some problems in the application process. Considering the above application problems, a new epoxy vitrimer system with special 3R properties (reprocessable, reparable and recyclable) is chosen in this work, in comparison to other vitrimer materials where no catalyst is needed to activate the 3R properties under high temperature.

Considering the impact of flame retardants on environment, more and more eco-friendly flame retardants had been studied widely to polymers aiming to improve the flame retardancy. [20-26] Amongst these flame-retardant studies, DOPO and its derivatives are one of the common phosphorus flame retardants, and have been used in epoxy system widely. A new DOPO derivative of DOPMPA synthesized by DOPO and piperazine was used as flame retardant of EP. Until the addition of flame retardant was 13 *wt*%, EP/DOPMPA blend passed a UL-94 V-0 rating and possessed lower release of heat and smoke[27]. Furthermore, Shanglin Jin et al. also reported flame retardant ABD of EP formed by acrolein and DOPO. Only adding 3 *wt*% of ABD can make EP acquire a high limit oxygen index and level of UL-94[28]. Phosphorus compounds show a good influence on improving flame retardancy of epoxy. DOPO and its derivatives play a role in not only gas phase but also condensed phase because of reactive P-H bond. Then phosphorus-containing compounds would be a good choice to act as flame retardant for epoxy.

Both flame retardancy and recyclability are important properties for epoxy resins. Herein, this work aims to fabricate a kind of recyclable flame-retardant epoxy, keeping balance of excellent flame retardancy and recyclability. Friendly DOPO is used as flame retardant of recyclable epoxy resin. The flame retardancy of epoxy improves by exploiting the free radical trapping mechanism in gas phase and carbon formation mechanism in condensed phase. Even if a small amount of flame retardant is adding, epoxy can gain effective flame-retardant performance. Furthermore, flame retardant has excellent compatibility with matrix, and flame-retardant epoxy still displays good transparency. In addition, mechanical properties of composites do not exhibit reduction. A new kind of recyclable flame-retardant epoxy has been synthesized, which can provide guidance to develop other novel recyclable flame-retardant epoxy resin.

## 2. Results and discussion

### 2.1 Reactivity between REP and DOPO

The P-H bond in DOPO has a certain activity because of the adjacent strong electron-absorbing group P=O bond. The epoxy group has a high degree of tension because of its three-membered ring, which is active and easy to be opened. The two are easier to react to form phosphorus-containing epoxy composites, the corresponding structure is shown in **Figure 1. (A)**. Both of DSC and FTIR tests are carried out to detect the reaction between epoxy group and reactive P-H bond. For better showing the reaction, the weight ration of REP and DOPO is around 10:3 for DSC and FTIR tests.

DSC curves **(Figure 1. (B))** recorded the thermal convert process during the reaction of epoxy group with P-H bond. The heat phenomenon of REP does not appear any change during the whole heating course. However pure DOPO shows a sharp peak at about 118 °C, corresponding to the melting endothermic process. Cured REP sample also does not display intensive heat, meaning the curing process finished completely. But there is a gently endothermic peak at preheating temperature of around 90 °C of EP-30% DOPO sample, demonstrating the reaction between epoxy group and P-H bond[29].

FTIR test was also used for further verifying the pre-reaction between epoxy group and P-H bond of DOPO. Just like the FTIR curves shown in **Figure 1. (C)**, both of raw materials epoxy and DOPO appear their characteristic peaks. For REP sample, the peaks are located at 1250 cm$^{-1}$, 1510 cm$^{-1}$, 2980 cm$^{-1}$, 915

cm$^{-1}$ and 3397 cm$^{-1}$ respectively, corresponding to C-O stretch vibration peak, benzene ring stretching vibration peak, alkyl stretching peak, the characteristic absorption peak of oxirane group and hydroxyl group. In the spectrum of DOPO, the peaks are located at 1230 cm$^{-1}$, 1590 cm$^{-1}$, 2430 cm$^{-1}$ and 3360 cm$^{-1}$ respectively, corresponding to the vibration of P=O bond, the Ph skeleton, the distinct P-H bond and hydrogen stretching vibration of Ph skeleton[30-32]. However, the EP-30% DOPO composite displays the different characteristic peak in comparison to REP and DOPO. The hydroxyl group shifts to 3510 cm$^{-1}$ from the peak at 3430 cm$^{-1}$ of REP sample because of chemistry environment change around hydroxyl group. That indicates some EP has reacted with DOPO, causing chemistry environment change. In addition, by compared with the Tg of REP, Tg values of REP composites with different flame retardant DOPO decreases gradually with the increase of DOPO content (**Figure 1. (D)**). This might cause by the reduction of cross-linking density of epoxy thermosets, being also attributed to the reaction between P-H bond and epoxy group. These results show the pre-reaction between epoxy and flame retardant, causing the well mixture. This will play an important role in improving properties of epoxy, especially the transparency **(Figure S5)**.

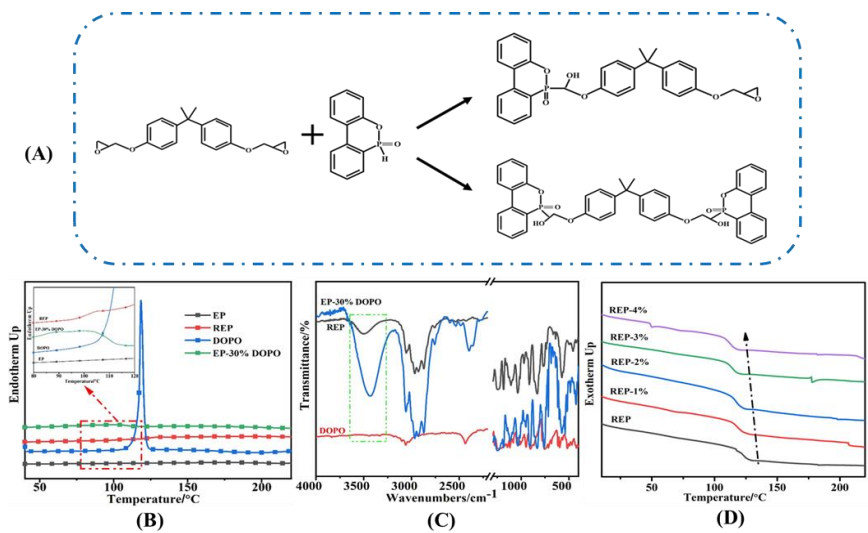

Figure 1. (A): Schematic outline of phosphorus-containing epoxy resin[33]; DSC curves (B) and FTIR spectrum (C) of different samples; (D) DSC curves of REP composites.

## 2.2 Fire behaviors of REP composites

As small-scale fire tests, the LOI and UL-94 tests were applied to evaluate the flammability of REP composites first of all. Neat REP with a LOI value of 21.7% is flammable material and difficult to self-extinguish. After ignition of pure REP, it can be burned out into a flash immediately, accompanied by the formation of a large amount of smoke and dripping fiercely which can ignite cotton. It also fails to pass UL-94 level. Once adding flame retardant DOPO, both the LOI values and UL-94 level of REP composites improve. As the amount of DOPO raises from 1 *wt*% to 4 *wt*%, the LOI value of REP composites gradually increases from 24.3 % to 29.5 %. In the meanwhile, REP composites containing DOPO present dramatically better self-extinguish ability after removing ignition, even at low loading of DOPO. Though REP-1% blend does not self-extinguished, its dripping after burning does not ignite the cotton in comparison to pure REP. Along with the content of DOPO reached 2 *wt*%, REP-2% shows a significant effect in UL-94 test (V-2). In addition, both of REP-3% and REP-4% composites can reach V-0 level at UL-94 test with a total burning time less than 10 s. The LOI and UL-94 results demonstrate that DOPO can show the obvious influence on improving flame retardancy. This is mainly due to the free radical scavenging capability showing by phosphoric containing flame-retardant DOPO, which can suppress flame combustion process.

CCT was carried out to further evaluate combustion performance of REP composites seriously. The parameters include pHRR, total heat release (THR), effective heat combustion (EHC) and residue weight, corresponding curves over time and data are displayed in **Figure 2** and **Table S2**. For pure REP specimen, it holds a sharp HRR peak with the value of 1369 kW/m$^2$. After adding DOPO, the curves undergo an evident decline gradually over test time. The pHRR values of REP composites cut down to 974 kW/m$^2$, 835 kW/m$^2$, 870 kW/m$^2$ and 774 kW/m$^2$ respectively. This indicates the flame retardant can improve flame retardancy significantly of REP system. In addition, it is noteworthy that REP composites containing DOPO all present two small slope peaks. The formation and accumulation of protective char layer caused in the early stage can be acted as barrier for hindering the spread of heat and oxygen. Once the accumulation of heat and volatiles reaches a certain extent, the initial char layer might be broken and further fire grow to form second peak. Furthermore, the pHRR of REP composites becomes gentle gradually, which means the formation of the stronger of char layer resulting from the catalytic action of phosphorus from DOPO, especially by comparison to pHRR of pure REP. Meanwhile, THR of REP composites also show decrease trend compared with pure REP. EHC$^e$ values of REP composites also decrease with the increase of DOPO, implying that DOPO displays strong quenching effect and combustion inhibition effect in gas phase. The time-varying trend of residual mass of REP composites is similar to TGA curves, which further shows the promoted carbonization by DOPO. Both of these results demonstrate better flame retardancy owning the exist of phosphorus, leading to the improvement of catalytic charring capabilities and free radical scavengers' effect in gas phase.

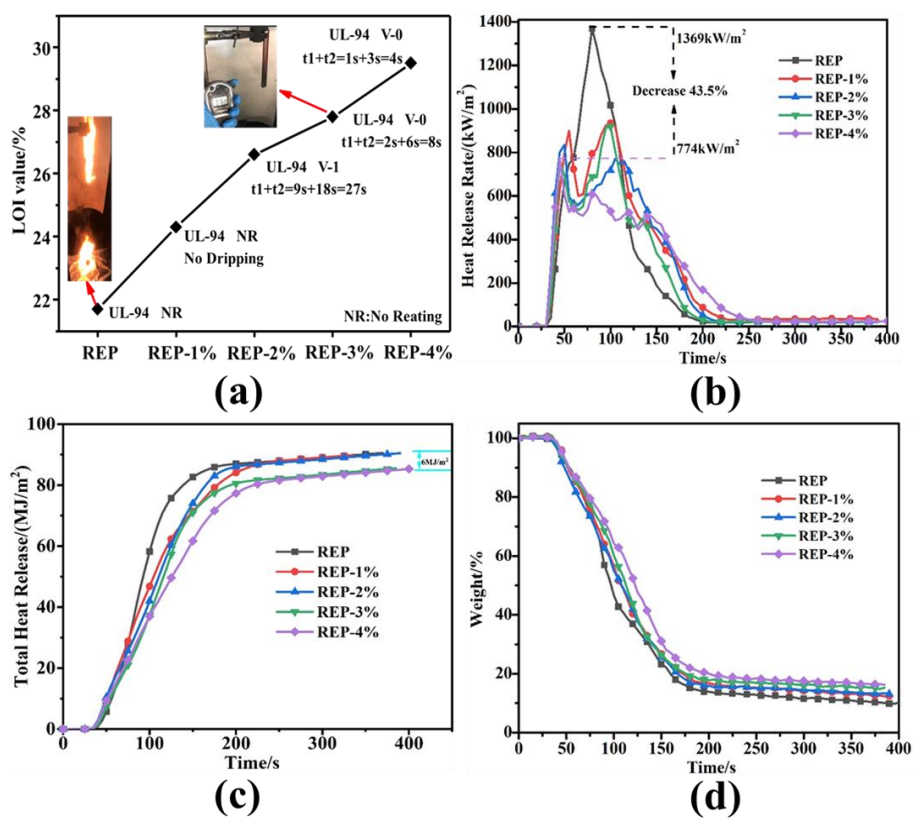

Figure 2. (a) the summary of LOI and UL-94 tests results; (b) HRR, (c) THR, (d) residue weight vs time curves of REP composites.

## 2.3 Flame retardant mechanism

In an effect to investigate the condensed phase mechanism, SEM analysis and Raman spectra were carried out to detect different microstructure of residue char of REP composites. **Figure 3. (A)** displays the digital and SEM images of different residue char after CCT test. Pure REP burns out completely and there are few residual chars left. So, the structure of residue char of REP is porous and fragmentary, that means the char layer is not dense enough to inhibit the escape of decomposed gases. However, once adding DOPO into REP composites, the char morphologies show an obvious discrimination. As the amount of DOPO increases, the char structure gradually becomes more compact and integrated, which can improve the intumescent effect. This indicates that addition of DOPO can form flame shield during combustion.

The microstructure of residue chars was further analyzed by Raman spectrum. As shown by curves in **Figure 3. (B)**, all curves appear one broad peaks at 1352 cm$^{-1}$ and one sharp peak at 1594 cm$^{-1}$ representatively, belonging to D and G peak respectively. Usually, D peak is associated with vibrations of carbon atoms with dangling bonds, G peak is associated with vibration in all sp$_2$ bonded carbon atoms in a 2-dimensional hexagonal lattice[34,35]. Moreover, the intensity ratio of D peak to G peak is calculated to evaluate residual char's graphitization degree. As seen, in comparison to $I_D/I_G$ value of pure REP, $I_D/I_G$ value of REP composites increases with the increase of flame retardant, displaying the smaller size of carbonaceous microstructure[36]. That means the formation of better char layer of REP composites, which can be served as a better protective barrier. It is beneficial to improve flame retardancy of REP system.

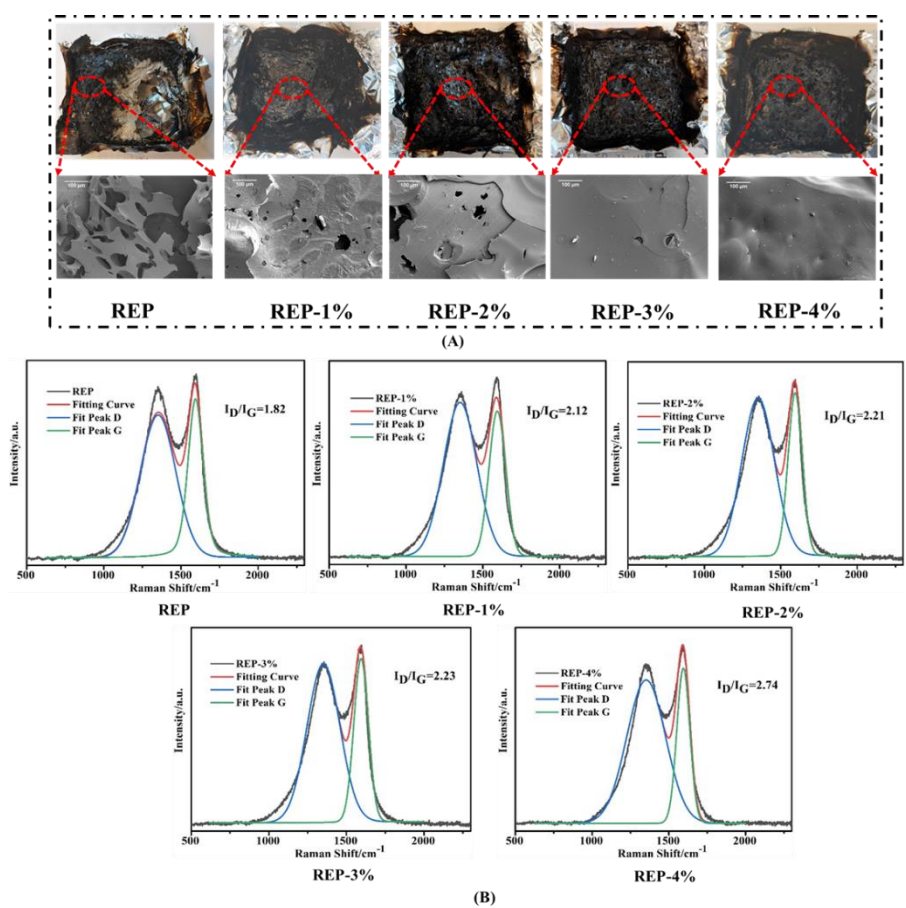

Figure 3. (A) Digital images of residue char and corresponding SEM images of residual char for REP composites; (B) Raman spectra of residual char for REP composites.

According to the above discussion results, the possible flame-retardant mechanism is proposed and the corresponding scheme is shown as below. In this recyclable flame-retardant REP system, DOPO displays a positive role in gas phase and condensed phase because of the excellent compatibility between small amount of DOPO and epoxy matrix. In the condensed phase, REP composites containing P element can facilitate the dehydration and char formation, showing better barrier to inhibit the release of heat and oxygen. Meanwhile, REP composites can produce some P-containing free radical such as PO· and

HPO· free radical to quench H· and OH· active radicals caused from combustion, which can further terminate the combustion chain reaction in gas phase. Both of the positive effect in the condensed phase and free radical scavenging in gas phase caused by flame retardant lead to superb flame retardancy.

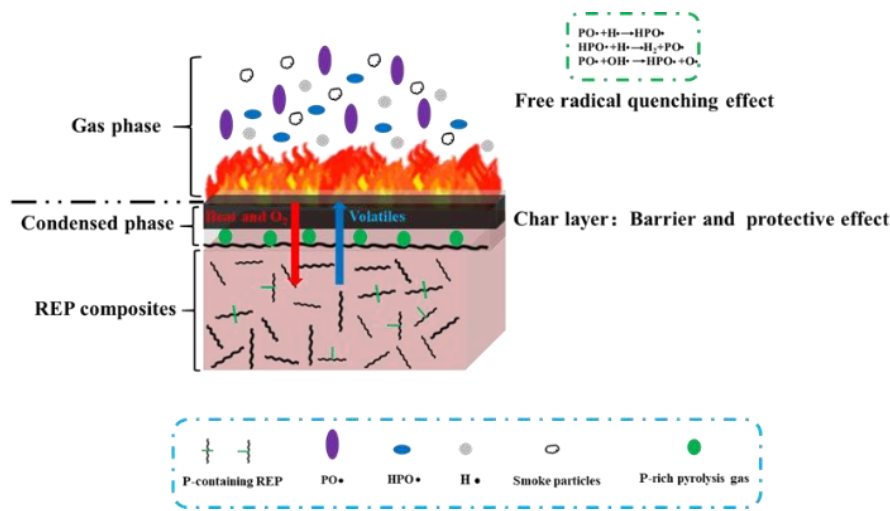

Figure 4. Schematic illustration of combustion mechanism.

## 2.4 Mechanical behaviors of REP composites

In order to check if the addition of flame retardant can affect mechanical and dynamic properties of flame-retardant epoxy vitrimers, both stress relaxation measurements tensile tests were carried out. A 1 % of strain is applied at different temperatures above the Tg and the relaxation modulus over time was monitored. The relaxation time ($\tau^*$) at each temperature is determined as the time needed to relax the 67 % of the initial relaxation modulus $G_0$ according to Maxwell model for viscoelastic fluids. The temperature dependence of $\tau^*$ for all vitrimers is fit to an Arrhenius relationship so as to calculate the activation energy (Ea) of flame-retardant epoxy vitrimers. All Arrhenius plots present a linear correlation of ln ($\tau$) with 1000/T in the measured temperature range.

Normalized stress relaxation curve of REP composites at different temperatures is shown in **Figure S2**. The activation energy of REP composites calculated is in line with previously obtained results[16, 17], are166 KJ/mol, 152 KJ/mol, 136 KJ/mol and 123 KJ/mol respectively. A slight decrease is due to the consumption of epoxy groups with DOPO, which can reduce the crosslinking density of formulation with the increase of DOPO content. It is observed that the lower of Tg causes the shorter of relaxation time at the same temperature. This observation is attributed to the enhanced diffusion and exchange of reactive groups due to its lower crosslinking density and viscosity. On the other hand, there is a decrease on the activation energy. That is because of the more thermal dependence of the exchange kinetics of the networks when increasing the Tg. Moreover, the DMA results are also shown in **Figure S3**. Compared with different REP composites, there is also no evident decrease in these results.

## *2.5 Recyclability of REP composites*

The presence of dynamic covalent bonds in the flame-retardant REP vitrimers permits the recycling of this thermoset materials since the dynamic covalent bonds are capable to exchange by heating above its Tg. In an attempt to investigate the recyclability of flame-retardant epoxy vitrimers, just REP-3% sample was chosen to verify. Pictures in **Figure 5** show detailed process. Cured REP-3% was grinded into powder firstly and the obtained powder was hot-pressed at 200 ºC at 100 bars for 5 min. It is easy to see that defect free laminate is obtained after reprocessing process. Moreover, REP-3% composite still expresses relatively high transparency even adding flame retardant.

To fully characterize the original and the recycled samples, they are subjected to uniaxial tensile test until break, Dumbbell shape specimens of the pristine and recycled REP-3% are tested with a universal tensile machine **(Figure S4)**. The reprocessed REP-3% composite shows almost the same mechanical properties as pristine material (90 % of initial tensile strength is achieved). Recycled REP-3% sample displays higher disparity than the pristine material in the tensile strength values due to presence of porosity in the lamina due to lack of pressure in some zones. Theoretically, this process could be repeated as many times as desired, although repeated processing might cause the aging of the material leading to lower mechanical properties.

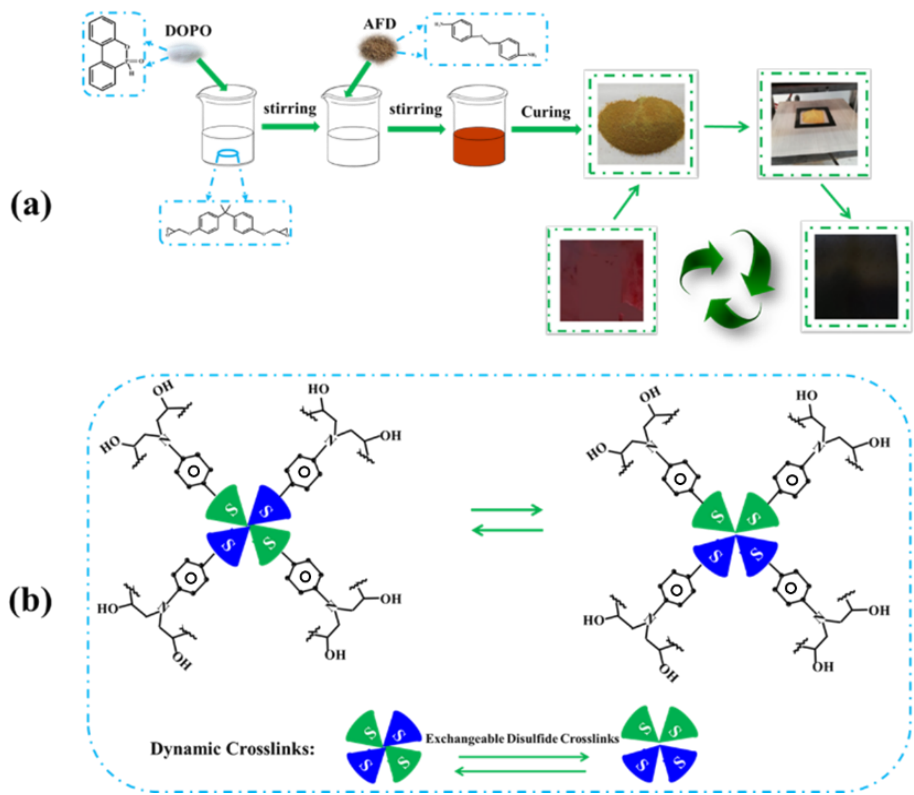

Figure 5. (a) the detailed recyclability process of REP-3%; (b) schematic diagram of exchangeable disulfide crosslinks.

## 3. Conclusion

The facile way was used to prepare recyclable flame-retardant epoxy in this work. The flame retardancy of REP composites is significantly improved even at the low loading of flame retardant. By comparison to neat REP sample, REP-3% composites only adding 3 % DOPO into epoxy can pass V-0 level of UL-94 test, with the relatively high LOI value of 27.8 %. In addition, pHRR of REP-4% decreases by about 43.5 % to the value of 774 kW/m$^2$, which attributes to accelerated charring ability to inhibit oxygen. The CCT results also show a decrease of EHC$^e$ value, indicating the DOPO acts in gas phase by free radicals scavenging effect. It is worth nothing that epoxy can still display recyclability even if the addition of flame retardant because of exchangeable disulfide crosslinks. Furthermore, REP composites show good transparency and same mechanical properties with pure REP sample. This research paves the way to obtain recyclable flame-retardant epoxy materials. It is also more conducive to broaden the application of epoxy.

## Acknowledgements

This work is partly funded by China Scholarship Council, China under the Grant CSC (201908110272).

## Funding Sources

This work did not receive any specific funding from funding agencies in public, commercial.